\documentclass[conference]{IEEEtran}
\usepackage[english]{babel}

\usepackage[letterpaper,top=2cm,bottom=2cm,left=3cm,right=3cm,marginparwidth=1.75cm]{geometry}

\usepackage{xurl} 

\usepackage{amsmath}
\usepackage{graphicx}
\usepackage[colorlinks=true, allcolors=blue]{hyperref}

\begin{document}

\title{IID-Based QPP-RNG: A Random Number Generator Utilizing Random Permutation Sorting Driven by System Jitter}

\author{
    \IEEEauthorblockN{Randy Kuang, Dafu Lou}
    \IEEEauthorblockA{
        Quantropi Inc. \\
        1545 Carling Ave., Suite 620\\
        Ottawa, Ontario, K1Z 8P9\\
        Emails: \{randy.kuang, dafu.lou\}@quantropi.com
        }
}

\maketitle

\begin{abstract}
We propose a groundbreaking random number generator that achieves truly uniform, independent, and identically distributed (IID) randomness by integrating Quantum Permutation Pads (QPP) with system jitter--derived entropy, herein called IID-based QPP-RNG. Unlike conventional RNGs that use raw timing variations, our design uses system jitter solely to generate ephemeral QPP pads and derives 8-bit outputs directly from permutation counts, eliminating the need for post-processing. This approach leverages the factorial complexity of permutation sorting to systematically accumulate entropy from dynamic hardware interactions, ensuring non-deterministic outputs even from fixed seeds. Notably, IID-based QPP-RNG achieves a min-entropy of 7.85-7.95 bits per byte from IID min-entropy estimate, surpassing ID Quantique's QRNG (7.157042 bits per byte), which marks a breakthrough in randomness quality.
Our implementation employs a dynamic seed evolution protocol that continuously refreshes the internal state with unpredictable system jitter, effectively decoupling the QPP sequence from the initial seed. Cross-platform validation on macOS (x86 and ARM) and Windows (x86) confirms uniformly distributed outputs, while evaluations compliant with NIST SP 800-90B show a Shannon entropy of 7.9999 bits per byte. Overall, IID-based QPP-RNG represents a significant advancement in random number generation, offering a scalable, system-based, software-only, post-quantum secure solution for a wide range of cryptographic applications.
\end{abstract}

\begin{IEEEkeywords}
QPP, OTP, Random Number Generator, RNG, PRNG, TRNG, Post-Quantum Cryptography, system Jitter, Uniform Distribution
\end{IEEEkeywords}

\section{Introduction}

Random number generation (RNG) is a fundamental component of modern cryptography, secure communications, and statistical modeling. The reliability of cryptographic systems depends on high-entropy randomness; failures in RNG design can lead to severe security vulnerabilities, as demonstrated by numerous attacks exploiting predictable randomness \cite{NIST-SP800-90A}. Existing RNGs include pseudo-random number generators (PRNGs), which rely on deterministic algorithms, hardware random number generators (HRNGs), which leverage physical entropy sources, and system-level entropy sources that exploit computational nondeterminism. Each approach presents distinct trade-offs in security, complexity, and efficiency.

Modern RNG methodologies have evolved along three principal axes. Pseudo-random number generators (PRNGs), such as linear congruential generators \cite{Lehmer1949} and the Mersenne Twister \cite{Matsumoto1998}, offer deterministic efficiency but require external entropy sources for cryptographic security. Cryptographically secure PRNGs (CSPRNGs) improve resilience using hash-based \cite{Ryan2022-csprng} and block cipher-based \cite{block-cipher-csprng} constructions but remain susceptible to quantum attacks \cite{nist_pqc2023}. Permutation-based approaches, including Fisher-Yates shuffling \cite{permutation-prng-FisherYates1938}, introduce alternative entropy mechanisms through combinatorial randomness, forming the basis for our QPP-RNG design.

Hardware-based solutions leverage thermal noise \cite{IntelRNG2012}, avalanche effects \cite{hrng-2023, hrng-2011-wang}, quantum optical processes \cite{IDQuantique2020}, and vacuum fluctuations \cite{Gabriel2010}. While offering strong security properties, these methods often suffer from cost and implementation complexity \cite{Ma2016,Bierhorst2018}. System-level RNGs, such as Müller’s timing jitter RNG \cite{muller_cpu_jitter} and the Linux kernel entropy pool, utilize microarchitectural noise but face challenges related to side-channel vulnerabilities in virtualized environments \cite{kerrigan2012study,vano2020eboot}.

The emergence of quantum computing poses significant threats to classical cryptographic schemes. Algorithms such as Shor’s \cite{shor1994} and Grover’s \cite{Grover1996} have demonstrated the ability to compromise widely used encryption methods. As a result, governments and industry leaders, particularly in financial and defense sectors, are actively preparing for post-quantum security challenges.

To address these concerns, the Quantum Permutation Pad (QPP) was introduced as a quantum-secure symmetric cryptographic framework. QPP utilizes high-entropy quantum permutation spaces to achieve bijective transformations, offering a strong foundation for cryptographic security \cite{kuang2020shannon, qpp-springer-kuang-2022}. Unlike classical Boolean logic, where an $n$-bit system has an information space size of $2^n$, the corresponding quantum permutation space contains $2^n!$ quantum permutation operators. This factorial expansion results in an effective Shannon information entropy of $\log_2(2^n!)$, making it highly resistant to brute-force attacks.

Building on QPP principles, we present IID-Based QPP-RNG, a novel random number generator that derives randomness from permutation counts rather than direct system jitter variations. Unlike traditional approaches that extract entropy from system timing jitter, this method ensures independent and identically distributed (IID) randomness by counting the number of permutations required for sorting a randomly permuted array and mapping the count modulo 256. This approach enhances statistical independence and eliminates non-IID biases typically associated with system jitter measurements.

Our results demonstrate that IID-Based QPP-RNG provides a scalable, software-only solution for cryptographic-grade randomness generation, establishing a new paradigm for post-quantum secure RNGs.

\section{IID-Based QPP-RNG}

The Quantum Permutation Pad (QPP), introduced by Kuang and Bettenburg \cite{kuang2020shannon} and further developed by Kuang and Barbeau \cite{qpp-springer-kuang-2022}, extends Shannon’s concept of perfect secrecy (i.e., the One-Time Pad) into the quantum domain by leveraging the inherent non-commutativity of permutation operations. This property allows for repeated pad usage without compromising security—a significant advantage over conventional OTP, which requires single-use pads.

QPP has since evolved into a versatile cryptographic framework. Kuang et al. have applied QPP to pseudo-random number generation \cite{kuang2021pseudo}, symmetric and asymmetric cryptography \cite{qpp-springer-kuang-2022, kuang2021quantum1, kuang2021quantum2, kuang2023-hppk-kem, kuang2024-hppk-ds}, and quantum computing implementations using quantum permutation gates \cite{kuang2022-epj-qpp, perepechaenko2023-epj-qpp}. A recent review \cite{kuang2025qpp} provides a comprehensive overview of QPP's role in quantum-secure schemes.

In our IID-Based QPP-RNG, random numbers are derived from permutation counts modulo 256, ensuring that the output is independent and identically distributed (IID). Rather than using raw system jitter as a direct randomness source, our method integrates QPP-based permutation sorting with system jitter measurements to update the seed of an LCG-based PRNG. This guarantees that each QPP pad generated for sorting is random and unpredictable, and that the resulting permutation count—when reduced modulo 256—produces an IID 8-bit output. In effect, system jitter indirectly drives the process by providing the unpredictable QPP pads used to sort the disordered array, while the output solely depends on the permutation counts.

For example, consider an ordered array
\[
A = \{0, 1, 2, 3, 4, 5\}
\]
or its corresponding computational basis
\[
\{|0\rangle, |1\rangle, |2\rangle, |3\rangle, |4\rangle, |5\rangle\}.
\]
With a secret permutation \(P\), the array \(A\) is permuted into
\[
C = \{5, 3, 1, 0, 4, 2\}
\]
(or its equivalent cipher basis). To recover \(A\), the sorting algorithm attempts to determine the inverse permutation \(P^{-1}\) by repeatedly applying random permutations until the array is sorted. The number of permutation attempts, denoted \(N\), is recorded and then reduced modulo 256 to yield an 8-bit random value:
\[
d = N \bmod{256}.
\]
Although a 6-element array has a permutation space of \(6! = 720\) possibilities (approximately 9.4 bits of entropy), the key randomness stems from the variability in \(N\), which is influenced by the system jitter derived from the QPP pad.

System jitter represents the inherent, unpredictable timing variations in a computing system—arising from cache behavior, branch prediction, pipeline fluctuations, and OS scheduling. While direct extraction of system jitter can lead to non-IID outputs, our approach leverages jitter indirectly to drive the LCG PRNG, thereby generating unpredictable QPP pads for each sorting operation. This variability in pad selection translates into randomness in the permutation count, which, when mapped modulo 256, produces a statistically IID 8-bit output.

The IID-Based QPP-RNG begins with an initialization phase using a static seed, which is subsequently decoupled by reseeding with 64 bits of system jitter-derived randomness. During the generation phase, the QPP-GEN (a PRNG such as an LCG) produces pseudo-random numbers that are employed by the Fisher-Yates shuffling algorithm to generate a permutation \(P_i\). The sorting algorithm applies \(P_i\) to a disordered array and checks for a sorted order. The number of permutation attempts is recorded, and upon successful sorting, this count is reduced modulo 256 to produce an 8-bit random output. Additionally, the sorting time delta—also reduced modulo 256—is used to update the LCG seed (e.g., by left-shifting the current seed by 8 bits and adding the reduced time delta), ensuring that the QPP pad for the next sorting operation remains unpredictable.

This process is iteratively repeated, ensuring that each random output is statistically independent through an evolved dynamic seed per sorting. Note that the method for extracting timing variations may vary across operating systems based on their clock resolutions. For instance, on Windows one might use \(r_i = t_i/100 \bmod{256}\); on macOS (x86) \(r_i = t_i \bmod{256}\); and on macOS (ARM) \(r_i = t_i/10 \bmod{256}\).

\begin{figure}[ht]
    \centering
    \includegraphics[scale=0.4]{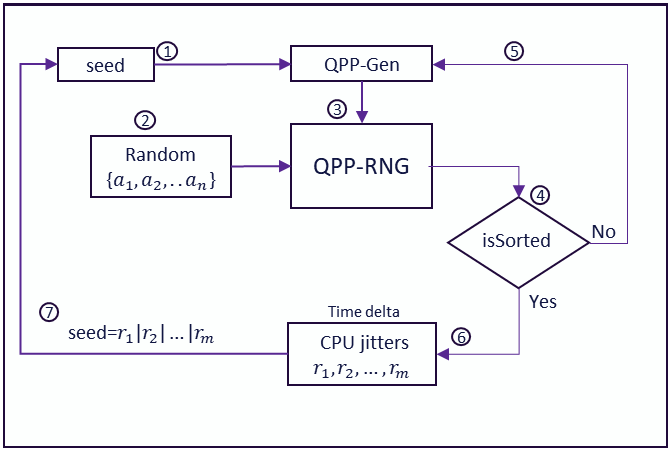}
    \caption{QPP-RNG Initialization: This phase eliminates dependence on deterministic seeds by incorporating system jitter into the seeding process.}
    \label{fig:init}
\end{figure}

\begin{figure}[ht]
    \centering
    \includegraphics[scale=0.4]{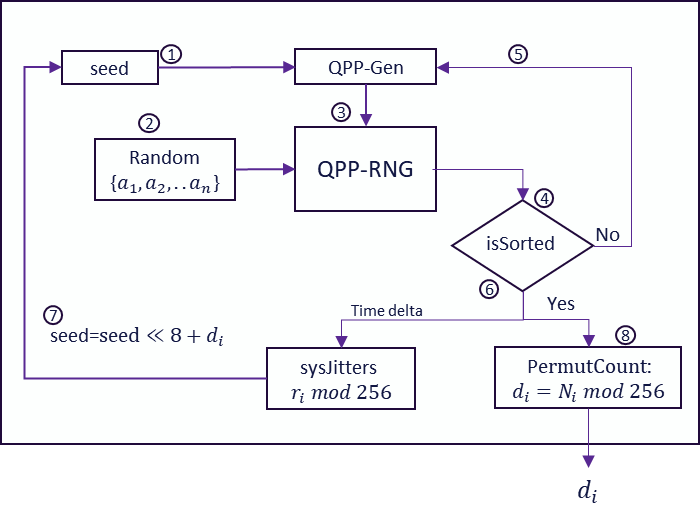}
    \caption{QPP-RNG Random Number Generation: This phase generates the final random numbers and updates the seed for the next iteration based on the permutation count and timing information.}
    \label{fig:rng}
\end{figure}

Figures~\ref{fig:init} and \ref{fig:rng} illustrate the initialization and random number generation phases, respectively, highlighting how system jitter influences the permutation process and contributes to the IID random outputs.

IID-Based QPP-RNG thus provides a robust, software-only solution for generating cryptographically secure, IID random numbers. By combining the exponential entropy of quantum permutation spaces with the inherent unpredictability of system jitter—transformed via permutation counts modulo 256—our method offers a scalable, quantum-resistant RNG suitable for modern cryptographic applications.

\section{Implementation and Testing}

To evaluate the performance and randomness quality of IID-Based QPP-RNG, we implemented the algorithm in Java and tested it on three representative systems:
\begin{itemize}
    \item \textbf{macOS (x86):} Running macOS Ventura 13.x or Sonoma 14.x on an Intel Core i7/i9 MacBook Pro.
    \item \textbf{macOS (ARM):} Running macOS Ventura 13.x on an Apple M1 Pro MacBook Pro, with a clock resolution of 10 ns (ARMv8.5 cycle counter).
    \item \textbf{Windows (x86):} Running Windows 11 22H2 on a 12th Gen Intel Core i5-1240P processor, with a clock resolution of 100 ns.
\end{itemize}
Our experiments focused on analyzing how system jitter influences the permutation sorting process and the resulting permutation count, which is used to derive the final 8-bit random outputs (via modulo 256).

\begin{figure}[ht]
    \centering
    \includegraphics[scale=0.45]{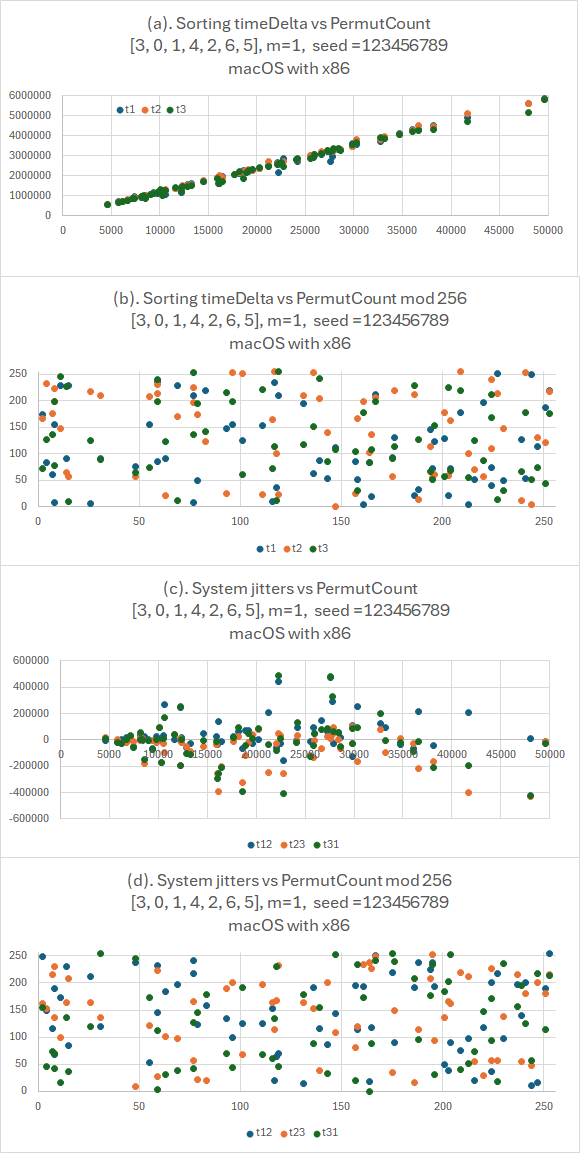}
    \caption{Sorting time vs. permutation counts on macOS (x86). The upper graph shows the actual system time delta (ns) against permutation counts for a disordered array of 7 elements, while the lower graph presents the same data reduced modulo 256. Data points labeled $t_1$, $t_2$, and $t_3$ represent three consecutive runs using a fixed seed (123456789).}
    \label{fig:macOS_x86}
\end{figure}

For initial illustration, we conducted experiments using fixed QPP pads generated pseudo-randomly from an LCG seeded with 123456789—temporarily omitting the full initialization phase described in Fig.~\ref{fig:init}. This allowed us to compare the distributions of system time deltas and permutation counts. Figure~\ref{fig:macOS_x86} shows sorting time variations as a function of permutation counts on a macOS (x86) system.

Figure~\ref{fig:macOS_x86}(a) illustrates that for a 7-element array (with approximately 12.3 bits of entropy), the average sorting time is roughly proportional to the number of permutation attempts. However, even at the same permutation count, noticeable fluctuations in the time delta occur due to system jitter. These fluctuations arise from several factors, including:
\begin{itemize}
    \item \textbf{Memory Access Patterns:} Fisher-Yates shuffling involves frequent array accesses, and cache misses can force slower main memory retrieval.
    \item \textbf{TLB Misses:} Delays from Translation Lookaside Buffer (TLB) misses contribute additional variability.
    \item \textbf{Memory Controller Contention:} Simultaneous memory requests from multiple system components (e.g., GPU, other processes) introduce further delays.
\end{itemize}
Thus, the time delta \(t_i\) for a sorting operation can be expressed as:
\[
t_i = \bar{t}_i \pm \Delta_i,
\]
where \(\bar{t}_i\) is the average execution time for the given number of permutations and \(\Delta_i\) represents the total system jitter. Figure~\ref{fig:macOS_x86}(c) highlights these fluctuations by subtracting successive time delta distributions.

Because the Shannon entropy from the permutation space of a 7-element array exceeds 8 bits, we can extract 8 bits of randomness by reducing either the time delta or the permutation count modulo 256, as shown in Figures~\ref{fig:macOS_x86}(b) and \ref{fig:macOS_x86}(d).

Our experiments also reveal significant variability in the actual number of permutations required for sorting. Although the average for a 7-element array is \(7! = 5040\), observed counts ranged from below this average to as high as 50,000. This wide range illustrates that the number of permutations required to recover the original order is unbounded in the worst case, and this variability is crucial for mitigating biases in QPP-RNG.

To fully decouple the deterministic QPP sequence from the fixed LCG seed, we include an initialization phase. In this phase, eight independent sorting operations on a disordered array yield eight 8-bit time delta values that are concatenated to form a 64-bit seed used to reseed the LCG. This process ensures that subsequent QPP pad generation is driven by an evolving, entropy-rich seed.

During each sorting operation, we record two key variables: the system time delta,
\[
\text{sysJitter} \bmod 256 \rightarrow r,
\]
which captures 8 bits of jitter, and the permutation count,
\[
N \mod 256 \rightarrow d,
\]
which represents an 8-bit random value derived from the number of permutations executed to complete sorting. The inherent non-determinism of system jitter ensures that the seed used to generate QPP pads remains unpredictable and varies over time and across systems. We incorporate \(\text{sysJitter}\) into the seed evolution using the following update rule:
\begin{equation}
    \text{seed} = (\text{seed} \ll 8) + \text{sysJitter}.
\end{equation}
This update guarantees that each sorting operation is governed by a QPP pad generated from an evolving, entropy-rich seed. Consequently, each generated 8-bit output (derived from the permutation count modulo 256) benefits from an entropy source exceeding 8 bits, thereby enhancing overall randomness quality.

\begin{figure}[ht]
    \centering
    \includegraphics[scale=0.3]{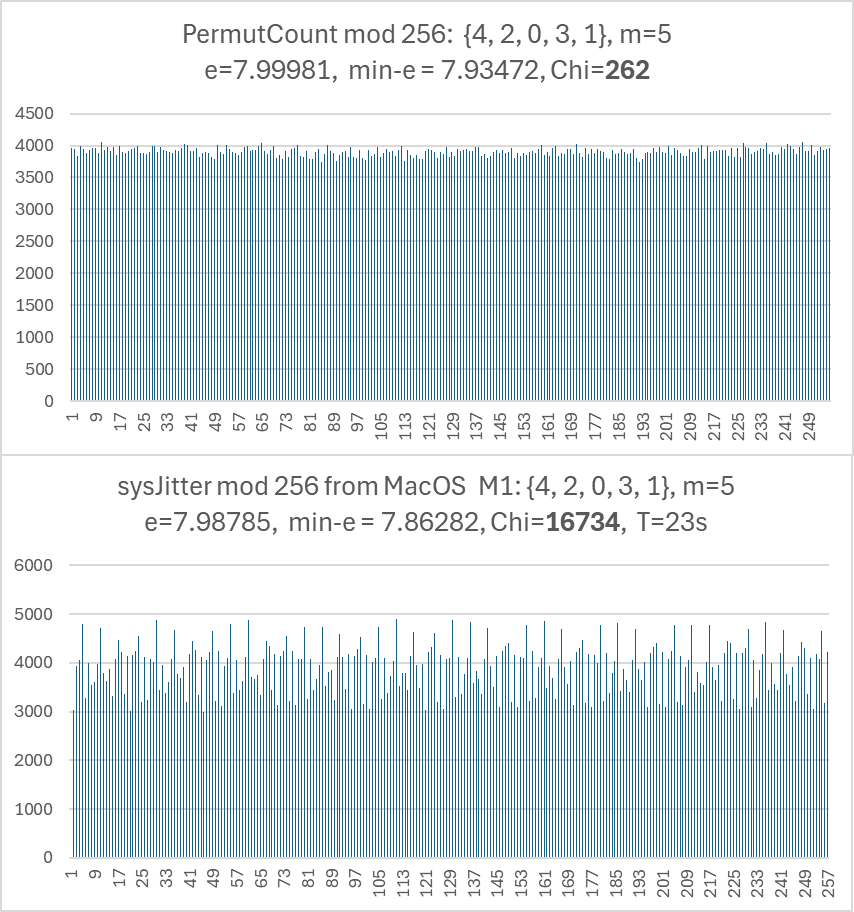}
    \caption{Random number distributions from QPP-RNG time deltas (mod 256) on macOS (ARM). A corresponding permutation count distribution (mod 256) is shown for reference. The input array consists of 5 elements.}
    \label{fig:random-m54}
\end{figure}

\begin{figure}[ht]
    \centering
    \includegraphics[scale=0.32]{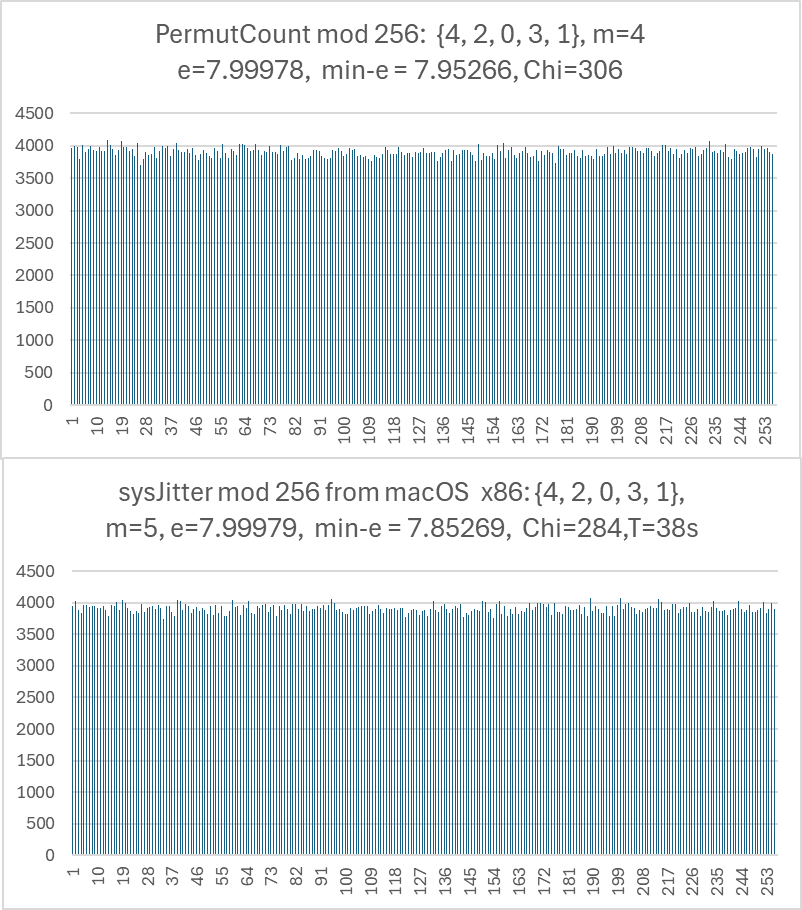}
    \caption{Random number distributions from QPP-RNG time deltas (mod 256) on macOS (x86). A corresponding permutation count distribution (mod 256) is shown for reference. The input array consists of 5 elements.}
    \label{fig:random-macOS}
\end{figure}

\begin{figure}[ht]
    \centering
    \includegraphics[scale=0.35]{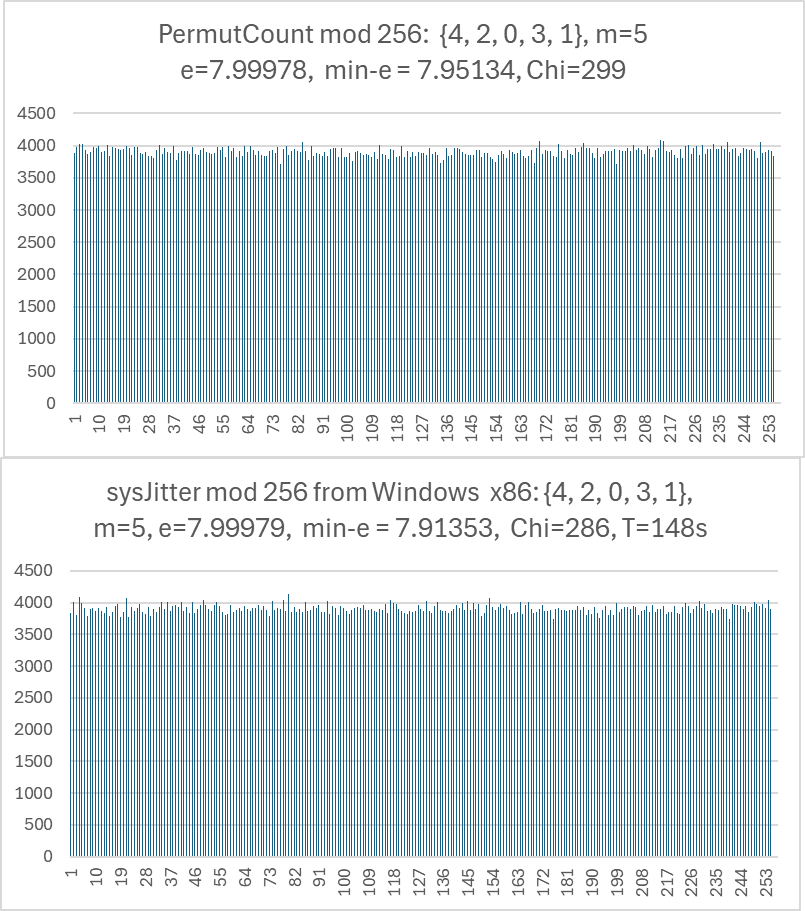}
    \caption{Random number distributions from QPP-RNG time deltas (mod 256) on Windows (x86). A corresponding permutation count distribution (mod 256) is shown for reference. The input array consists of 5 elements.}
    \label{fig:random-windows}
\end{figure}

To further assess the statistical properties of the IID random outputs, we analyzed the frequency distributions of 8-bit numbers derived from both time deltas and permutation counts (each reduced modulo 256) using a 5-element input array. Figures~\ref{fig:random-m54}, \ref{fig:random-macOS}, and \ref{fig:random-windows} present these distributions for macOS (ARM), macOS (x86), and Windows (x86), respectively.

For a 5-element array, which provides approximately 6.9 bits of entropy, at least \(m=4\) repetitions are necessary to extract a full 8-bit random output. We computed the \textbf{Shannon entropy} and \textbf{min-entropy estimate for IID} in accordance with NIST SP 800-90B. For example, repeating the sorting 5 times yielded an effective entropy of 9.23 bits. In one test on macOS (ARM) in Figure~\ref{fig:random-m54}, the permutation count distribution achieved a Shannon entropy of 7.99981 bits per byte and a min-entropy of 7.93472 bits per byte for IID min-entropy estimate, comparing with the min-entropy 7.154 bit/byte from IDQ QRNG~\cite{cameron-randomness-2024}, with a Chi-square value close to 262 (near the ideal of 256). In contrast, the raw system jitter distribution exhibited slightly lower Shannon entropy (7.98785 bits) and min-entropy estimate 7.86282 bits/byte with a much higher Chi-square value (16734), indicating greater bias. This bias can be mitigated by either increasing the repetition \(m=7\) or array size to 6 or 7. 

Figures~\ref{fig:random-macOS} and \ref{fig:random-windows} further demonstrate that on macOS (x86) and Windows (x86), both the permutation count and system jitter distributions yield high-quality randomness, although the Windows implementation operated at approximately 3.5 times slower speed than the macOS (x86) version.

In summary, our experiments—conducted over 1,000,000 QPP-RNG outputs across three system configurations—demonstrate that the permutation count distribution consistently achieves high entropy and uniformity, confirming the IID properties of our approach. The combination of quantum permutation pad entropy with system jitter–driven seed evolution yields a robust, software-only RNG that is scalable and quantum-resistant, making it well-suited for modern cryptographic applications. In practical deployments, the output random numbers are derived solely from the permutation count, ensuring that QPP-RNG functions as a true IID RNG.

\section{Security Analysis of QPP-RNG}

IID-Based QPP-RNG is designed for cryptographic-grade randomness by combining dynamically generated Quantum Permutation Pads (QPP) with system jitter–derived entropy to produce IID random numbers from permutation counts. Its security rests on two pillars: (1) the post-quantum security of ephemeral QPP pads and (2) the unpredictability of permutation counts obtained during the sorting process. The secrecy of the permutation sequences underpins QPP’s post-quantum security, and since QPP pads are never reused—each is generated from an evolving seed that is initially derived from system jitter—the approach decouples any fixed seed from future operations. During random number generation, the seed is continuously updated with fresh system jitter entropy, and an 8-bit value derived from the permutation count is used to form the final random output. Consequently, even if a previous QPP pad or permutation count were compromised, subsequent outputs would remain secure.

A primary security challenge is the reliable extraction of system jitter entropy for seed evolution. Although system jitter originates from non-deterministic system interactions (e.g., memory contention, branch prediction delays, interrupts), its role here is to drive the generation of QPP pads rather than serving directly as the randomness source. The permutation counts—obtained during the sorting process and reduced modulo 256—are the final random numbers, and their unpredictability is indirectly ensured by the quality of the jitter-derived seed. Therefore, careful entropy estimation and conditioning are essential to maintain high security.

The seed evolution protocol mitigates state predictability risks by continuously refreshing the internal state with both jitter measurements and permutation counts. Although a theoretical adversary might try to influence timing behavior (e.g., through controlled cache contention), such attacks require deep system knowledge and fine-grained control, rendering them impractical under typical operating conditions.

Platform-specific clock resolution also impacts security. Timing granularity varies significantly (e.g., 1 ns on macOS x86, 10 ns on macOS ARM, and 100 ns on Windows x86). Lower resolution can amplify quantization effects in observed jitter, complicating entropy estimation. QPP-RNG adapts to these differences by adjusting array sizes or increasing repetition counts to ensure sufficient entropy is captured for reliable seed evolution.

In summary, IID-Based QPP-RNG achieves robust security by combining the post-quantum secrecy of ephemeral QPP pads with the unpredictability of permutation counts derived from the sorting process. Its quantum resistance stems from the infeasibility of reconstructing ephemeral permutation sequences, while dynamic seed evolution—fueled by system jitter—ensures forward secrecy. 

\medskip
\noindent
\textbf{Future Directions:} Further work will focus on formal security proofs and comprehensive side-channel analyses. Recent advances in quantum machine learning, as discussed in \cite{facial-quantum-ml}, have underscored the critical role of high-quality randomness in quantum state preparation. The IID-based QPP-RNG, with its breakthrough min-entropy of 7.85-7.95 bits per byte (surpassing ID Quantique's QRNG at 7.157042 bits per byte), provides a robust, scalable, and post-quantum secure source of randomness. Integrating such randomness into quantum machine learning pipelines, for example in facial identification systems, may further enhance the resilience and reliability of these emerging applications.

\section{Conclusion}

This paper introduced IID-based QPP-RNG, a novel random number generator that combines Quantum Permutation Pads (QPP) with system jitter entropy to achieve independent and identically distributed (IID) randomness. IID-based QPP-RNG models sorting as a decryption process, leveraging QPP and the factorial time complexity of permutation-based sorting to accumulate entropy from system-level interactions. Crucially, IID-based QPP-RNG guarantees non-deterministic output, even with a fixed initial seed, leveraging system jitter's inherent unpredictability. Notably, it achieves uniform randomness without post-processing, setting a new benchmark for system-level IID RNGs.

Rigorous cross-platform testing (macOS x86/ARM, Windows x86) validated the IID properties of QPP-RNG. System jitter-based time deltas, scaled to clock resolution and reduced modulo 256, consistently produced uniform 8-bit outputs. A dynamic seed evolution protocol, generating a 64-bit seed from system jitter and iteratively refreshing it using permutation counts, enhances security. Empirical evaluations (NIST 800-90B compliant) demonstrated exceptional statistical quality (Shannon entropy $\geq 7.999$ bits/byte, min-entropy $\geq 7.85-7.95$ bits/byte), confirming cryptographic-grade randomness. Small platform-dependent jitter variations (particularly on macOS ARM) had a negligible impact on the output random numbers derived from permutation counts.

In summary, IID-based QPP-RNG offers a post-quantum-secure, hardware-agnostic solution that uniquely bridges algorithmic complexity (QPP) with system-level entropy (system jitter). By deriving randomness solely from permutation counts, IID-based QPP-RNG ensures that each output is independent and identically distributed, distinguishing it from conventional RNGs. This capability, combined with its resilience to quantum attacks and adaptability to diverse architectures, positions IID-based QPP-RNG as a compelling alternative, particularly in environments lacking dedicated entropy sources. Future work will explore performance optimizations, extended statistical evaluations, and integration into cryptographic protocols.

\bibliographystyle{ieeetr}
\bibliography{my}

\begin{thebibliography}{10}

\bibitem{NIST-SP800-90A}
E.~Barker and J.~Kelsey, ``{NIST Special Publication 800-90A Rev. 1: Recommendation for Random Number Generation Using Deterministic Random Bit Generators},'' Tech. Rep. SP 800-90A Rev. 1, National Institute of Standards and Technology (NIST), June 2015.

\bibitem{Lehmer1949}
D.~H. Lehmer, ``Mathematical methods in large-scale computing units,'' {\em Annals of the Computation Laboratory of Harvard University}, vol.~26, pp.~141--146, 1949.

\bibitem{Matsumoto1998}
M.~Matsumoto and T.~Nishimura, ``Mersenne twister: a 623-dimensionally equidistributed uniform pseudo-random number generator,'' {\em ACM Trans. Model. Comput. Simul.}, vol.~8, pp.~3--30, 1998.
\newblock \url{https://api.semanticscholar.org/CorpusID:3332028}.

\bibitem{Ryan2022-csprng}
C.~Ryan, M.~Kshirsagar, G.~Vaidya, A.~Cunningham, and R.~Sivaraman, ``Design of a cryptographically secure pseudo random number generator with grammatical evolution,'' {\em Scientific Reports}, vol.~12, no.~1, p.~8602, 2022.

\bibitem{block-cipher-csprng}
C.~Petit, F.-X. Standaert, O.~Pereira, T.~G. Malkin, and M.~Yung, ``A block cipher based pseudorandom number generator secure against side-channel key recovery,'' in {\em Proceedings of the 2008 ACM Symposium on Information, Computer and Communications Security}, (New York, NY, USA), pp.~56 -- 65, Association for Computing Machinery, 2008.

\bibitem{nist_pqc2023}
{National Institute of Standards and Technology (NIST)}, ``Post-quantum cryptography standardization,'' 2023.
\newblock \url{https://csrc.nist.gov/projects/post-quantum-cryptography}. Ongoing project for standardizing quantum-resistant cryptographic algorithms.

\bibitem{permutation-prng-FisherYates1938}
R.~A. Fisher and F.~Yates, {\em Statistical Tables for Biological, Agricultural and Medical Research}.
\newblock Edinburgh, UK: Oliver Boyd, 1938.

\bibitem{IntelRNG2012}
{Intel Corporation}, ``Intel digital random number generator (drng) software implementation guide,'' tech. rep., Intel Corporation, 2012.
\newblock \url{https://software.intel.com/en-us/articles/intel-digital-random-number-generator-drng-software-implementation-guide}.

\bibitem{hrng-2023}
G.~Guerrer, ``Rava: An open hardware true random number generator based on avalanche noise,'' {\em IEEE Access}, vol.~11, pp.~119568--119583, 2023.

\bibitem{hrng-2011-wang}
F.-X. Wang, C.~Wang, W.~Chen, S.~Wang, F.~Lv, D.-Y. He, Z.-Q. Yin, H.~Li, G.~Guo, and Z.~Han, ``Robust quantum random number generator based on avalanche photodiodes,'' {\em Journal of Lightwave Technology}, vol.~33, pp.~3319--3326, 2015.
\newblock \url{https://api.semanticscholar.org/CorpusID:38355948}.

\bibitem{IDQuantique2020}
{ID Quantique}, ``Quantum random number generation: A cryptographic edge,'' 2020.
\newblock \url{{https://www.idquantique.com/random-number-generation/overview/}}.Accessed: 2025-02-05.

\bibitem{Gabriel2010}
C.~Gabriel, C.~Wittmann, D.~Sych, R.~Dong, W.~Mauerer, U.~L. Andersen, C.~Marquardt, and G.~Leuchs, ``A generator for unique quantum random numbers based on vacuum states,'' {\em Nature Photonics}, vol.~4, pp.~711--715, 2010.

\bibitem{Ma2016}
X.~Ma, X.~Yuan, Z.~Cao, {\em et~al.}, ``Quantum random number generation,'' {\em npj Quantum Information}, vol.~2, p.~16021, 2016.

\bibitem{Bierhorst2018}
P.~Bierhorst, E.~Knill, S.~Glancy, Y.~Zhang, A.~Mink, S.~Jordan, Y.-K. Liu, B.~G. Christensen, S.~W. Nam, M.~J. Stevens, and L.~K. Shalm, ``Experimentally generated randomness certified by the impossibility of superluminal signals,'' {\em Nature}, vol.~556, p.~223–226, 2018.

\bibitem{muller_cpu_jitter}
S.~Müller, ``Cpu jitter based non-physical true random number generator,'' tech. rep., chronox, 2014.
\newblock \url{https://www.chronox.de/jent/CPU-Jitter-NPTRNG-v2.2.0.pdf}, Accessed: Feb 6, 2025.

\bibitem{kerrigan2012study}
B.~Kerrigan and Y.~Chen, ``A study of entropy sources in cloud computers: Random number generation on cloud hosts,'' in {\em Computer Network Security}, vol.~7531 of {\em Lecture Notes in Computer Science}, (Berlin, Heidelberg), pp.~286--298, Springer, 2012.

\bibitem{vano2020eboot}
F.~Vaño-García and H.~Marco-Gisbert, ``E-boot: Preventing boot-time entropy starvation in cloud systems,'' {\em IEEE Access}, vol.~8, pp.~61872--61890, 2020.

\bibitem{shor1994}
P.~Shor, ``Algorithms for quantum computation: discrete logarithms and factoring,'' in {\em Proceedings 35th Annual Symposium on Foundations of Computer Science}, pp.~124--134, 1994.

\bibitem{Grover1996}
L.~K. Grover, ``A fast quantum mechanical algorithm for database search,'' in {\em Proceedings of the Twenty-Eighth Annual ACM Symposium on Theory of Computing}, STOC '96, (New York, NY, USA), p.~212–219, Association for Computing Machinery, 1996.

\bibitem{kuang2020shannon}
R.~Kuang and N.~Bettenburg, ``Shannon perfect secrecy in a discrete hilbert space,'' in {\em 2020 IEEE International Conference on Quantum Computing and Engineering (QCE)}, pp.~249--255, IEEE, 2020.

\bibitem{qpp-springer-kuang-2022}
R.~Kuang and M.~Barbeau, ``Quantum permutation pad for universal quantum-safe cryptography,'' {\em Quantum Information Processing}, vol.~21, p.~211, 2022.

\bibitem{kuang2021pseudo}
R.~Kuang, D.~Lou, A.~He, C.~McKenzie, and M.~Redding, ``Pseudo quantum random number generator with quantum permutation pad,'' in {\em 2021 IEEE International Conference on Quantum Computing and Engineering (QCE)}, pp.~359--364, IEEE, 2021.

\bibitem{kuang2021quantum1}
R.~Kuang, D.~Lou, A.~He, and A.~Conlon, ``Quantum safe lightweight cryptography with {Quantum Permutation Pad},'' in {\em 2021 IEEE 6th International Conference on Computer and Communication Systems (ICCCS)}, pp.~790--795, IEEE, 2021.

\bibitem{kuang2021quantum2}
R.~Kuang, D.~Lou, A.~He, and A.~Conlon, ``Quantum safe lightweight cryptography with {Quantum Permutation Pad},'' {\em Advances in Science, Technology and Engineering Systems Journal}, vol.~6, pp.~401--405, 2021.

\bibitem{kuang2023-hppk-kem}
R.~Kuang and M.~Perepechaenko, ``Homomorphic polynomial public key encapsulation over two hidden rings for quantum-safe key encapsulation,'' {\em Quantum Information Processing}, vol.~22, p.~315, 2023.

\bibitem{kuang2024-hppk-ds}
R.~Kuang, M.~Perepechaenko, M.~Sayed, and D.~Lou, ``Homomorphic polynomial public key with the barrett transformation for digital signature,'' {\em Academia Quantum}, vol.~1, no.~1, 2024.

\bibitem{kuang2022-epj-qpp}
R.~Kuang and M.~Perepechaenko, ``Quantum encryption with quantum permutation pad in ibmq systems,'' {\em EPJ Quantum Technology}, vol.~9, p.~26, 2022.

\bibitem{perepechaenko2023-epj-qpp}
M.~Perepechaenko and R.~Kuang, ``Quantum encryption of superposition states with quantum permutation pad in ibm quantum computers,'' {\em EPJ Quantum Technology}, vol.~10, p.~7, 2023.

\bibitem{kuang2025qpp}
R.~Kuang, ``Quantum permutation pad for quantum secure symmetric and asymmetric cryptography,'' {\em Academia Quantum}, vol.~2, no.~1, 2025.

\bibitem{cameron-randomness-2024}
C.~Foreman, R.~Yeung, and F.~J. Curchod, ``Statistical testing of random number generators and their improvement using randomness extraction,'' {\em Entropy}, vol.~26, no.~12, 2024.

\bibitem{facial-quantum-ml}
P.~Easom-McCaldin, A.~Bouridane, A.~Belatreche, and R.~Jiang, ``Towards building a facial identification system using quantum machine learning techniques,'' {\em Journal of Advances in Information Technology}, vol.~13, pp.~198--202, April 2022.

\end{thebibliography}

\end{document}